\documentclass[sort&compress
    ,final            
  ]
  {aipproc}

\layoutstyle{8x11double}

\begin{document}

\title{Exploring Student Understanding of Energy through the Quantum Mechanics Conceptual Survey}


\author{S. B. McKagan}{
  address={Department of Physics and JILA, University of Colorado, Boulder, CO, 80309, USA}
}

\author{C. E. Wieman}{
  address={Department of Physics and JILA, University of Colorado, Boulder, CO, 80309, USA}
}

\begin{abstract}
We present a study of student understanding of energy in quantum mechanical tunneling and barrier penetration.  This paper will focus on student responses to two questions that were part of a test given in class to two modern physics classes and in individual interviews with 17 students.  The test, which we refer to as the Quantum Mechanics Conceptual Survey (QMCS), is being developed to measure student understanding of basic concepts in quantum mechanics.  In this paper we explore and clarify the previously reported misconception that reflection from a barrier is due to particles having a range of energies rather than wave properties.  We also confirm previous studies reporting the student misconception that energy is lost in tunneling, and report a misconception not previously reported, that potential energy diagrams shown in tunneling problems do not represent the potential energy of the particle itself.  The present work is part of a much larger study of student understanding of quantum mechanics.
\end{abstract}

\maketitle


\section{Introduction}
Quantum mechanics is a fascinating subject because it is so challenging to the intuition.  Learning quantum mechanics requires learning to accept such counterintuitive notions as ``particles'' reflecting off barriers even though they have enough energy to cross them as well as tunneling through barriers that they do not have enough energy to cross.  Perhaps even harder than accepting these notions is actually understanding them.  Previous research shows that even when students accept strange ideas, they often do not understand them \cite{Ambrose1999a,Bao1999a,Morgan2003a,Muller2005a}.  In order to change this, we must first gain a clearer understanding of how students actually think about these concepts.  We are in the process of developing an instrument called the Quantum Mechanics Conceptual Survey (QMCS) \cite{QMCS} to measure understanding of basic concepts in quantum mechanics.  The QMCS is a multiple-choice survey, designed to provide quantitative data to complement and extend the qualitative interview data that already exists on this subject.  Through in-class tests and student interviews, we have used the QMCS to elicit and explore student thinking about many concepts in quantum mechanics.  Here we focus on two QMCS questions that were developed to further explore student misconceptions presented in a previous study \cite{Ambrose1999a}.  We elaborate on the source and extent of these misconceptions and present a new misconception not seen in previous work.

We present results from two modern physics classes where the QMCS was given at the end of the Spring 2005 semester and from 17 student interviews.  The two classes used the same textbook \cite{Taylor2004a} and covered similar material, but were taught by different professors with different teaching styles.  One class was intended for engineering majors (ENGsp05, N=68) and one for physics majors (PHYSsp05, N=64).  The interview subjects included four students from ENGsp05, nine students from PHYsp05, and four students who took the equivalent of ENGsp05 in a previous semester (ENGfa04), taught by a different professor.  In interviews, students were asked to work through the QMCS, thinking out loud and explaining why they chose the answers they did.  The interviewer (SBM) asked questions to further probe their thinking.

There is some controversy in the Physics Education Research community over the definition of the word ``misconceptions'' and the extent to which students have them \cite{Elby2000a}.  In this paper we will take the perspective that student thinking can take many forms, ranging from fragmented and incoherent ideas that apply only in certain contexts to robust theories that are consistent across all relevant contexts.  Here we will use the word ``misconception'' to mean any incorrect student idea that can be clearly articulated and is seen consistently in numerous students in at least one context.

\section{Reflection: A range of energy?}

\begin{figure}[htbp]\label{barrier}
  \includegraphics[width=3.1in]{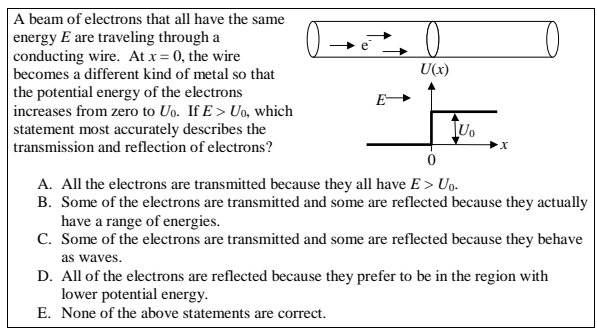}
  \caption{A barrier penetration question from the QMCS.  This question is adapted from an open-ended question in Ref. \cite{Ambrose1999a}, and the distracters are based on student responses reported therein.  We have changed the wording from ``Monoenergetic electrons'' to ``A beam of electrons that all have the same energy $E$'', and added the figure and description of the wire to make the question more grounded in physical reality.}
\end{figure}

In an extensive study of student understanding of wave properties of light and matter \cite{Ambrose1999a}, Ambrose has reported on the ``failure to recognize that reflection occurs at the boundary between regions of different potential or wave speed'' and the ``mistaken belief that reflection and transmission of a beam of particles is due to a range of energies of the particles in the beam.'' Using a survey with an open-ended question similar to that shown in Fig. \ref{barrier}, he found that many students did not believe that any electrons would be reflected, using the classical reasoning of answer A.  Of those students who knew that some electrons should be reflected, many thought the reason was that the beam contained electrons with a range of energies, as stated in answer B, in spite of the fact that the beam was described as ``monoenergetic''.  We adapted the question in Fig. \ref{barrier} from Ref. \cite{Ambrose1999a} in order to further explore these misconceptions.  We wanted to determine the extent to which students hold these misconceptions and why.  Further, we wanted to determine to what degree the misconception described by answer B was due to students' simple misunderstanding of the word ``monoenergetic,'' and whether it was robust enough to appear even if the contradiction between the problem statement and the answer was more apparent.  Our results show that this misconception is quite robust; most students hold on to it even when the contradiction is explicitly pointed out.

\begin{figure}[htbp]\label{q20results}
  \includegraphics[width=3.1in]{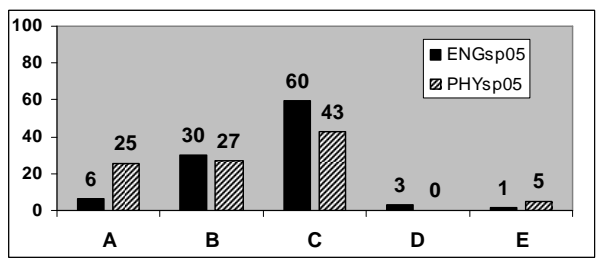}
  \caption{Percentage of students who selected each answer to the question shown in Fig. \ref{barrier} on a test given in two classes.}
\end{figure}

Fig. \ref{q20results} shows the answers that students selected for the question in Fig. \ref{barrier} on the in-class exam.  It should be pointed out that transmission and reflection through a barrier was discussed in both classes, and that all interview subjects from these two classes remembered that reflection occurred in some cases.  In ENGsp05 a nearly identical question had been discussed in class at great length, which probably explains why these students did better on this question than the PHYsp05 students, although the PHYsp05 students did better on most QMCS questions.  In spite of instruction, the test results, coupled with the interviews discussed below, show that a significant fraction of students in both classes held the misconceptions described by answers A and B.

Out of the 15 students interviewed on this question, eight students initially selected answer B.  In all of these cases, the interviewer then asked, ``How do you reconcile that answer with the statement in the question that all the electrons have the same energy?''  In response to this question, three students stuck by their answer, giving detailed justifications, two students changed their answers to C (the correct answer), two students changed to A, and one student changed to D.

Of the three students who defended answer B, two used the Heisenberg uncertainty principle, arguing that you could never really know the energy, and $E$ was just the average energy.  The third student gave a more elaborate explanation, based on a misunderstanding of a type of diagram commonly used quantum mechanics in which wave functions are drawn on top of energy levels: ``{\ldots}every picture I've ever seen where they tell us what the wave function is and they say it has this energy $E$, he draws a line down the middle and then draws the wave function around it.  And I guess I just internalize that as saying that\ldots that's like their average energy\ldots''

Of the 15 students interviewed, we argue that seven had a robust misconception that reflection at a barrier is caused by electrons having a range of energies.  In addition to the three students who defended answer B, there is strong evidence that the three students who switched to answers A or D, as well as one student who initially selected answer A, also held this misconception.  These four students all argued that in \textit{this} case, in which all the electrons had the same energy, they would all be transmitted (or reflected), but in all the other cases they had discussed in class, in which there was reflection, the electrons must have a range of energies.  We view this misconception as an extension of the first misconception discussed by Ambrose, that all electrons with sufficient energy should be transmitted.  It is essentially a way of reconciling the first misconception with the remembered fact that sometimes electrons are reflected.

\section{Tunneling: Energy loss and the meaning of Potential Energy}

Several previous studies have found that students often believe that particles lose energy in tunneling \cite{Ambrose1999a,Bao1999a,Morgan2003a,Muller2005a}.  While these studies provide extensive interview data on this misconception, there is little quantitative data on the extent to which it is held, as we provide here.

Morgan, Wittmann, and Thompson \cite{Morgan2003a} suggest several explanations for why students might believe that energy is lost in tunneling.  One explanation is that most textbooks and lecturers draw the energy and the wave function on the same graph, leading many students to confuse the two, believing that the energy, like the wave function, decays exponentially during tunneling.  A second explanation is classical intuition about objects physically passing through obstacles, in which energy usually is dissipated.  Muller and Sharma \cite{Muller2005a} propose another explanation: students may be thinking of the energy of an ensemble of particles, rather than the energy of a single particle.  In this case, since not all of the particles are transmitted, it is actually correct that an ensemble \textit{as a whole} loses energy during tunneling.

\begin{figure}[htbp]\label{energyloss}
  \includegraphics[width=3.1in]{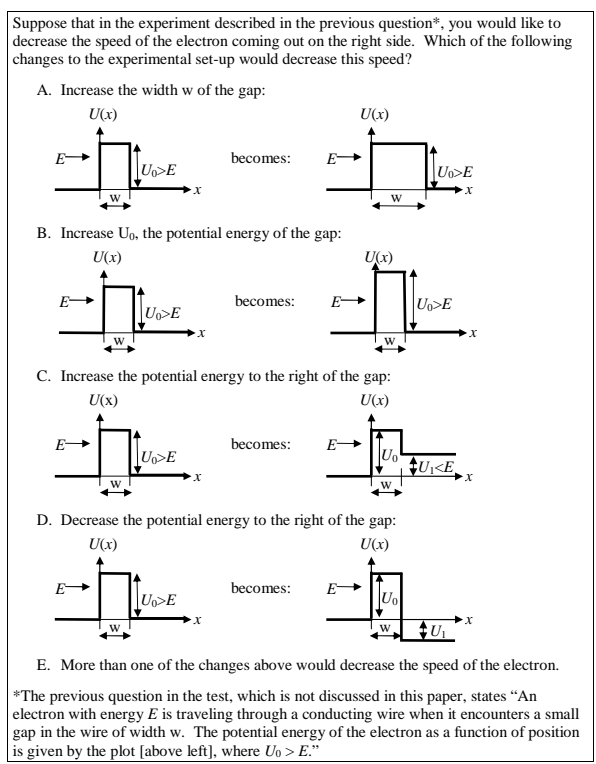}
  \caption{A tunneling question from the QMCS.  This question was developed to explore the misconception that energy is lost in tunneling.}
\end{figure}

\begin{figure}[htbp]\label{q10results}
  \includegraphics[width=3.1in]{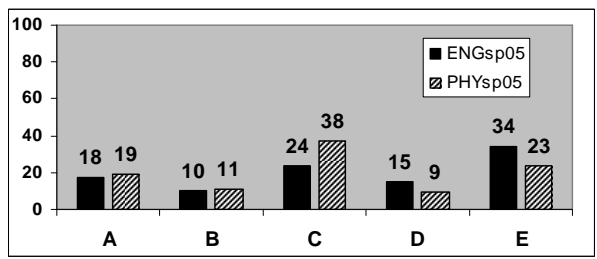}
  \caption{Percentage of students who selected each answer to the question shown in Fig. \ref{energyloss} on a test given in two classes.}
\end{figure}

Fig. \ref{energyloss} shows a QMCS question designed to elicit the misconception that energy is lost in tunneling.  In interviews, all students who selected answers A, B, or E argued that since energy was lost in tunneling, making the barrier wider and/or higher would lead to greater energy loss.  The fraction of students who selected one of these three answers (62\% in ENGsp05 and 53\% in PHYsp05)), shown in Fig. \ref{q10results}, gives us a lower bound on the fraction of students who hold the misconception that energy is lost in tunneling.  It is only a lower bound because in interviews, even students who gave the correct answer with the correct reasoning often second-guessed themselves and wondered whether energy might be lost in cases A and B as well.  Some students who initially held the misconception eventually chose the correct answer because cases D and E, which were unlike any examples they had seen in class, forced them to consider the energy on the right of the barrier more carefully. In interviews we saw evidence for all three of the explanations for this misconception listed above.

We found this question particularly useful in exploring student thinking about energy in tunneling, not only because it elicits the idea that energy is lost, but because understanding the correct answer, C, requires a clear understanding of the relationship of potential, kinetic, and total energy in the context of tunneling.  In many cases, this question elicited significant cognitive dissonance, as students struggled to reconcile two contradictory ideas: that energy is lost, and that kinetic plus potential equals total.  While it is technically possible to reconcile these ideas if it is the kinetic and total energy that is lost, we found that most students who thought that energy is lost did not have a clear idea of \emph{which} energy is lost.  When asked, they were just as likely to say potential energy as any other kind.  Often a single student would use two or even all three types of energy interchangeably within the same explanation.  Most of the interview subjects had fragments of both the correct view and the view that energy is lost simultaneously.

Of the four students interviewed from ENGsp05, \textit{all} held a robust misconception not seen in any previous study, and not seen in any of the interview subjects from the other courses.  These students thought that the quantity $U(x)$, plotted here and in nearly every problem involving solutions to the Schr\"{o}dinger equation, is \emph{not} the potential energy of the electron itself, but some kind of ``external energy.''  We discovered this misconception in the first interview conducted in this study, in which a student drew an exponentially decaying curve over the potential energy graph shown in the question, and consistently referred to this curve as representing ``the potential energy.''  The interviewer then asked what the graph shown in the question represented, since it was also referred to as ``the potential energy.''  The student replied, ``I don't know, that's just the bump that it goes through.  I don't know what it means.  I just see that and I know that it's some kind of obstacle that it goes through.''  When pressed, he said that the ``bump'' was ``the external energy that the electron interacts with'' and insisted that it was not the potential energy of the electron itself, in spite of the fact that it was explicitly labeled as such.  The interviews with other ENGsp05 students were very similar, with all of them referring to the graph as either ``the external potential energy'' or ``the potential energy of the medium,'' and quickly dismissing the idea that it was the potential energy of the electron itself.

It is unclear why this misconception was held so robustly by all of the interviewees from one course and not present at all in the interviewees from the other two courses.  The sample sizes are small and the courses, as well as the student populations, were different.  Therefore we do not wish to speculate on which factor caused the discrepancy.  However, it seems unlikely that this misconception is confined to students in this particular course, and we hope that other researchers will continue to probe student thinking about this topic in other contexts.

We find this misconception interesting because while many studies have shown that students think energy decays in tunneling, none of these studies discuss how students reconcile this idea with the fact that they are often drawing these decaying curves \textit{on top of} graphs of energy curves that are not decaying.

The source of this misconception becomes clearer if we consider that textbooks and lecturers nearly always refer to ``a particle in a potential'' as if the potential is something external.  One physics professor, in a discussion of these results with SBM, stated that the potential \emph{is} an external thing, caused by an external field of some kind.  SBM replied that it may be caused by an external field, but the potential energy is a property of the particle itself.  The professor said that he always thinks of it as a potential, rather than a potential energy, in which case it is not a property of the particle itself.  We suspect that many physics professors think this way, easily switching between potential and potential energy in their minds, forgetting that the distinction between the two is not a trivial factor of charge $q$ in the minds of their students.  In all other fields of physics, the symbol $V$ is used for potential and the symbol $U$ is used for potential energy, but most quantum mechanics textbooks use the symbol $V$ for potential energy (the textbook used in these classes is an exception to this rule) and then use the terms potential and potential energy interchangeably.  Few textbooks give any kind of physical explanation for the source of the potential energy term in the Schr\"{o}dinger equation.

Because we had seen this confusion over the meaning of potential through the semester, throughout the QMCS we used the symbol $U(x)$ rather than $V(x)$ and referred to this symbol as ``the potential energy function of the electron.''  In interviews students simply skimmed over this unfamiliar phrase and focused on the familiar symbols.

It is important to note that one should not over-interpret the statistical results given in Fig. \ref{q10results} as indicating the fraction of students in a class that hold a particular view.  First, answer E does not distinguish between students who think that A and B alone are correct and students who think that A, B, and C are correct.  Furthermore, even if this ambiguity were resolved, for example by allowing students to mark more than one correct answer, the answers alone would not tell us much about the thinking of those students who held fragments of the correct view and the view that energy is lost. In interviews we found that these students selected a wide range of answers.  Some students who held both views decided that A, B, and C were all correct, while others picked only one or two of these options, either at random or because one sounded slightly more plausible.  Several students changed their answer after several minutes of thinking through all the implications, which would not have happened under normal test-taking circumstances.

This project is the first step in a comprehensive study of student thinking about quantum mechanics.  We have found the QMCS to be a useful tool in probing student thinking, and will extend these results in further studies.

\begin{theacknowledgments}
  We would like to thank Dana Anderson and Mihaly Horanyi for useful feedback on the QMCS and for allowing us to give it to their classes, the physics education research group at the University of Colorado for additional feedback, Brad Ambrose, Michael Wittmann, Jeff Morgan, and Derek Muller for discussions of their studies on these topics, and finally, the students who participated in interviews.  This work was supported by the NSF.
\end{theacknowledgments}



\bibliographystyle{aipproc}   

\bibliography{PERCproceedings2005}

\begin{thebibliography}{7}
\expandafter\ifx\csname natexlab\endcsname\relax\def\natexlab#1{#1}\fi
\providecommand{\enquote}[1]{``#1''}
\expandafter\ifx\csname url\endcsname\relax
  \def\url#1{\texttt{#1}}\fi
\expandafter\ifx\csname urlprefix\endcsname\relax\def\urlprefix{URL }\fi
\providecommand{\eprint}[2][]{\url{#2}}

\bibitem[Ambrose(1999)]{Ambrose1999a}
B.~Ambrose, Ph.D. thesis, University of Washington (1999).

\bibitem[Bao(1999)]{Bao1999a}
L.~Bao, Ph.D. thesis, University of Maryland (1999).

\bibitem[Morgan et~al.(2003)]{Morgan2003a}
J.~T. Morgan, M.~C. Wittmann, and J.~R. Thompson, \emph{Phys. Educ. Res. Conf.
  Proceedings} (2003).

\bibitem[Muller and Sharma(2005)]{Muller2005a}
D.~A. Muller, and M.~D. Sharma, \emph{in preparation} (2005).

\bibitem[QMC(2005)]{QMCS}
http://cosmos.colorado.edu/phet/survey/qmcs/ (2005).

\bibitem[Taylor et~al.(2004)]{Taylor2004a}
J.~R. Taylor, C.~D. Zafiratos, and M.~A. Dubson, \emph{Modern Physics for
  Scientists and Engineers}, Pearson Prentice Hall, 2004, 2 edn.

\bibitem[Elby(2000)]{Elby2000a}
A.~Elby, \emph{J. Math. Behav.}, \textbf{19}, 481 (2000).

\end{thebibliography}

\IfFileExists{\jobname.bbl}{}
 {\typeout{}
  \typeout{******************************************}
  \typeout{** Please run "bibtex \jobname'" to optain}
  \typeout{** the bibliography and then re-run LaTeX}
  \typeout{** twice to fix the references!}
  \typeout{******************************************}
  \typeout{}
 }

\end{document}